\documentclass[twocolumn,showpacs,preprintnumbers,amssymb]{revtex4}

\usepackage{graphicx}% Include figure files
\usepackage{dcolumn}% Align table columns on decimal point
\usepackage{bm}% 
\usepackage{graphics}
\usepackage{epsfig}
\usepackage{hyperref}

\usepackage{graphics}
\usepackage{graphicx} 
 \usepackage{latexsym,amssymb}
\begin{document}
\preprint{APS/123-QED}
\title{\bf Modeling an efficient Brownian heat engine}

\author{Mesfin Asfaw}
\email{asfaw@cc.ncu.edu.tw}
\affiliation{Department of Physics and Graduate Institute of Biophysics\\
   National Central University, Jhongli, 32054, Taiwan}

\begin{abstract}

We discuss the effect of subdividing the ratchet potential on the performance of 
a tiny Brownian heat engine that is  modeled as a Brownian particle hopping in a viscous medium in a
sawtooth potential (with or without load) assisted by alternately  placed hot
and cold heat baths along its path. We show that 
 the   velocity, the efficiency and the coefficient of  performance of the refrigerator maximize when 
 the sawtooth potential is subdivided  into series of smaller connected barrier series.   When the engine operates quasistatically, 
 we  analytically  show that
  the efficiency of the engine can not approach the   Carnot efficiency and,   the coefficient of  performance of 
  the refrigerator is always less than the Carnot refrigerator due to the irreversible heat flow via the kinetic energy.
\end{abstract}
%\centerline{PACS. 05.40.Jc - Brownian motion}
%\centerline{PACS. 05.60.-k - Transport processes}
%\centerline{PACS. 05.70.-a - Thermodynamics}
\maketitle

 \section  { \bf Introduction}

Brownian heat  engine functions as transducer of thermal energy into mechanical work. It rectifies  thermal fluctuations into  a unidirectional current   as long as the system is out of equilibrium.  In the last few decades, the study of such a tiny  engine has got considerable attention  
not only  for the  construction of a miniaturized  engine that can help us to utilize  energy   resources at microscopic scales \cite{ph,r} but also for better understanding of the nonequilibrium statistical physics \cite{ken,gom}. The thermodynamic property of Brownian heat engine has been explored intensively  by considering  different  model systems. Brownian heat engine working due to spatially-variable temperature is one of  the model systems   which has been studied  in the pioneering works  by  B\"{u}ttiker \cite{butt87}, Van Kampen \cite{kampen88}, and Landauer \cite{land88}. After the  work of B\"{u}ttiker,    the  ratchet model has been the subject of several authors  \cite{miki3,Aus1,Ast2,bq,mesfin1,mesfin2}.  Recently, we considered an exactly solvable model of the heat engine and investigated the conditions under which  the model works as a heat engine, as a refrigerator and as neither of the two  \cite{mesfin3}. Not only we exposed the energetics of such an engine at a quasistatic limit but we also found the thermodynamic properties of the engine when it operates at a finite time.

In modeling of Brownian heat engine, one crucial but unexplored issue  is the way how to maximize the velocity, the efficiency and the coefficient of performance of the refrigerator. For instance, in construction of artificial Brownian  motors,  one may need  to design an engine that accomplishes its task  as fast as possible and efficiently. The  Brownian heat engine operates autonomously \cite{ken}. The performance of this engine (how fast and efficiently can it achieve its task?) relies upon the way how the ingredients of the system are arranged prior to  the engine operation.    The purpose of  this theoretical work is to present   one possible way of improving the performance of the heat engine. Earlier, the mean first passage time (MFPT) of a Brownian particle that walks over rugged sawtooth potential was studied by us \cite{mesfin4} and  in the work  \cite{m1,m2} using  super symmetric potential approach and using properties of random walk on networks formulated by Goldhirch and Gefen \cite{g1,g2}. The sawtooth potential was systematically subdivided into series of barriers  without altering the barrier height, the potential width and the area under the barrier.    The    theoretical works revealed the existence of an optimal barrier subdivision that minimizes the MFPT. In this work, we use similar approach   and study the effect of subdividing the ratchet potential on the performance of the engine.  We show that the velocity, the efficiency and the coefficient of performance of the refrigerator tend to increase  as  the ratchet potential is subdivided into series of barriers.

Unlike macroscopic heat engines, the Carnot efficiency is unattainable for Brownian heat engines when the engines work  quasistatically because of the  irreversible heat flow  via the kinetic energy \cite{ken,Aus1}. In this work, we obtain a simple analytic expression for the efficiency at quasistatic limit. The analytic result reveals that the efficiency of the engine never goes to the Carnot efficiency at quasistatic limit. Another important but unexplored issue is the  influence of the heat flow via the kinetic energy  on   the  coefficient of performance of the refrigerator. In the present work,    we analytically  show   that the coefficient of performance of the refrigerator is always less than the Carnot refrigerator when the engine operates  quasistatically.

The paper is organized as follows: In section II, we present the model. In section III, we study the dependence of the efficiency and the velocity on the model parameters in the absence of external force. 
 We show that  the velocity and the efficiency attain optimum values  at a particular value of barrier subdivision  $N$. At qausistatic limit, the efficiency never goes to Carnot efficiency for any $N$  as the heat transfer via the kinetic energy is irreversible.  In section IV, we consider the model in the presence of external load. We find that  the velocity, the efficiency and the coefficient of  performance of  the refrigerator attain maximum values when 
 the sawtooth potential is subdivided  into series of smaller connected barrier series.  We also show that  the efficiency of the engine never approaches the Carnot efficiency  and, the coefficient of performance of the refrigerator is always less than the  Carnot refrigerator at quasistatic limit.  Section V deals with summary and conclusion.

    \section{The model}

Consider a Brownian particle which walks in a viscous medium in a   periodic   sawtooth potential  whose potential profile (see Fig. 1) is described by
\begin{equation} U(x)=\cases{
   U_{0}[{x\over L_{0}}+1],&if $-L_{0}< x \le 0$;\cr
   U_{0}[{-x\over L_{0}}+1],&if $0 < x \le L_{0};$\cr
   }\end{equation} where $U_{0}$ and $L_{0}$ denote the barrier height and the width of the sawtooth potential, respectively.
The viscous medium is alternatively in contact with the hot $T_{h}$ and the  cold  $T_{c}$ reservoirs along the reaction coordinate as shown in Fig. 1. Using the same theoretical frame work \cite{m1},
the sawtooth potential is subdivided 
into series of smaller connected barrier series. For example, Fig. 2 shows the left and  the right sides  of the sawtooth potential  shown in Fig. 1 are subdivided into three small steps $N=3$.   
For single barrier step  between $x_{2}$ and $x_{4}$,  for simplicity, we 
choose  $U_{1}=2U_{2}$ and $a=2b$ where  $x_{4}-x_{2}=a+b$. In general, for $N$ equally spaced intervals, from the top of the barrier to either side, $L_{0}$ and $U_{0}$
are given by $L_{0}=Na+(N-1)b$ and $U_{0}=NU_{1}-(N-1)U_{2}$. Such parameterization is physically reasonable  as  the barrier height, the barrier width and  the area under the barrier remain approximately constant as $N$ is varied \cite{mesfin4,m1}.

\begin{figure}
\epsfig{file=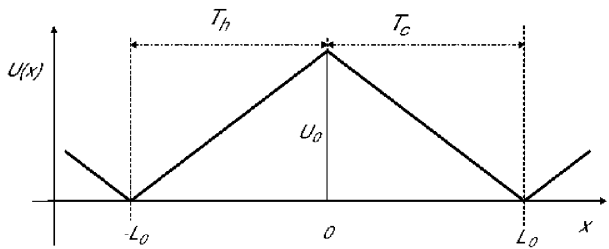,width=8cm}
\caption{Periodic sawtooth potential without load when the number of barrier subdivisions $N=1$. The sawtooth potential is  coupled with the hot $T_{h}$ and the cold  $T_{c}$ reservoirs. }
\end{figure}
\begin{figure}
\epsfig{file=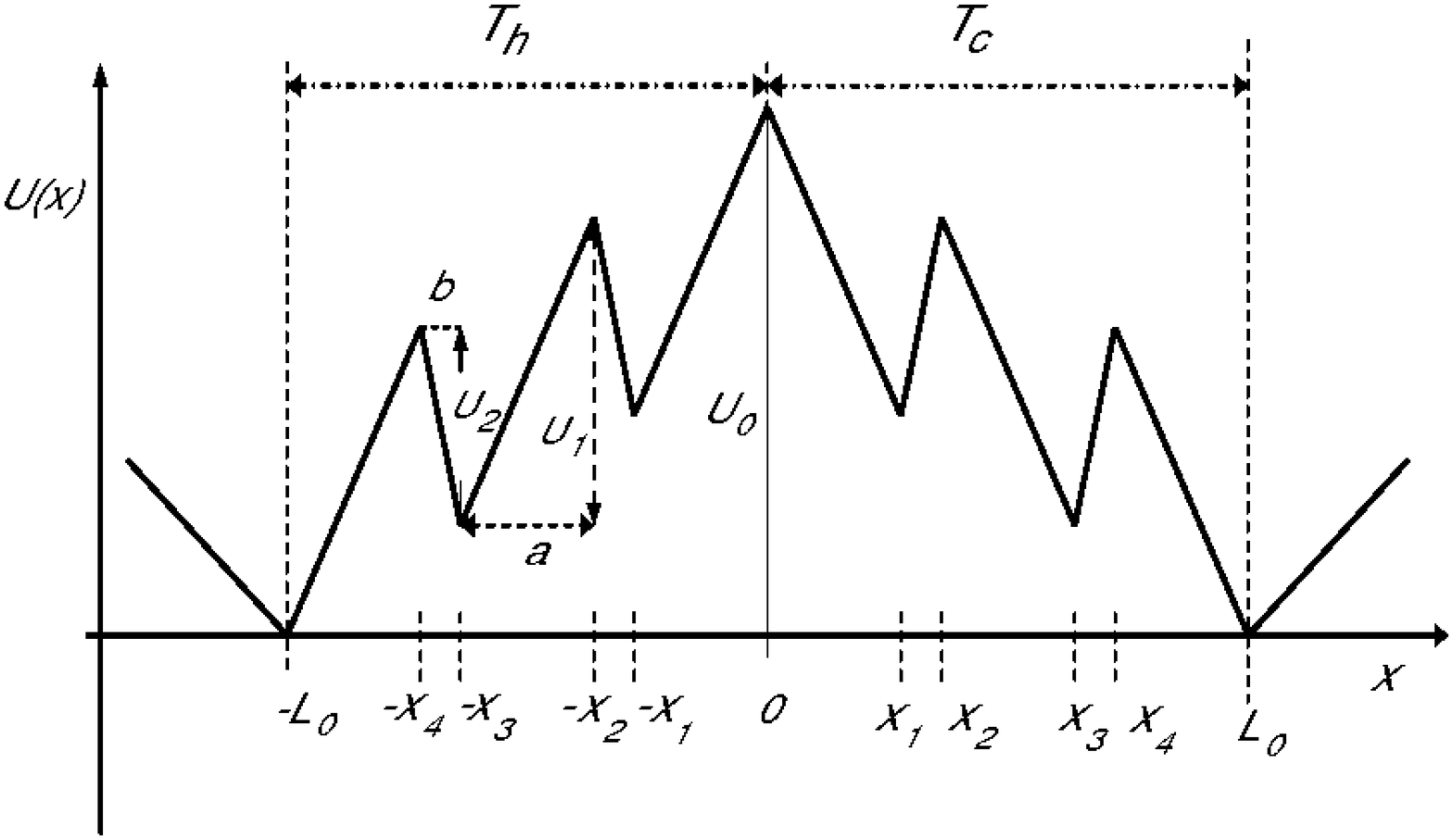,width=8cm}
\caption{Periodic rugged potential without load for $N=3$. The  potential is coupled with the hot $T_{h}$ and the cold  $T_{c}$ reservoirs. }
\end{figure}

The Brownian particle attains a directional motion when it is exposed to the potential coupled with spatially variable temperature. For such a  system, the general expression for the 
steady state current $J$ for the Brownian particle {\it in any periodic potential} with or without load is  reported in the work \cite{mesfin3}. The closed 
 form expression for the steady state current $J$ (please refer Appendix A,  ref. \cite{mesfin3}) is given by
\begin{equation}  J={-F\over G_{1}G_{2} +H F}.
 \end{equation}
 The drift velocity $v$ of the particle is associated to the steady state current $J$ and it is given by $v=2JL_{0}$.

 The hot reservoir is the ultimate source of energy for the  engine. When the engine works as a heat engine, the net flux of the particle is from hot to the cold heat baths. Hence when the particle moves from the hot to the cold heat baths, for any $N$, the particle takes 
$ U_{0}+v\gamma L_{0}+fL_{0}$ amount of energy from the hot reservoir  to surmount the potential of magnitude $U_{0}$ and  to overcome  the viscous drag force $v\gamma$ as well as  the external force  of amount $f$. On the other hand, $1/2k_{B}(T_{h}-T_{c})$ amount of energy is transferred  from   the hot to the cold heat baths via kinetic energy \cite{Aus1} when the particle walks from the hot to the cold heat baths.
 Hence the Brownian particle takes 
 \begin{equation}
Q_{h}=(U_{0}+fL_{0}+\gamma v L_{0})+{1\over 2}k_{B}(T_{h}-T_{c})
\end{equation}
amount of heat  from the hot reservoir. The  heat flow  to the cold reservoir is given by
\begin{equation}
  Q_{c}=U_{0}-L_{0}(f+\gamma v)+{1\over 2}k_{B}(T_{h}-T_{c}).
\end{equation}
When the engine acts as refrigerator,  the net flow of the particle is from the cold to the hot reservoirs. Note that, due to the particle recrossing between the hot and  the cold reservoirs, heat 
is leaking from the hot to the cold reservoir of magnitude $1/2 k_B (T_h - T_c)$.  This is in opposite direction to the heat being taken out of the cold reservoir. Hence, this quantity contributes as negative to $Q_c$. Thus,  the net heat flow out of   the cold reservoir is given by $Q_{c}=U_{0}-L_{0}(f+\gamma v)-{1\over 2}k_{B}(T_{h}-T_{c})$.

 Not all motors are designed to pull loads and alternative proposals for efficiency depend  on the task each motor performs. Some motors may have to achieve high velocity against a frictional drag. This basically implies that the objective of the motor is to move a certain distance in a given time interval. For such motors ($f=0$),  the useful work is   the difference between $Q_{h}$ and  $Q_{c}$:
$ W=Q_h- Q_c=2\gamma vL_{0}$.  For motors designed to pull loads  $f \ne 0$, the useful work is given by $W =2fL_{0}$. The efficiency $\eta$ and the coefficient of
performance (COP) of the refrigerator $P_{ref}$ of the engine  is given by
$ \eta=W/ Q_{h}$ and 
$ P_{ref} =  Q_{c}/ W$.

The purpose of this theoretical work, for given parameter values of 
 $U_{0}$ and $L_{0}$, to  find  the velocity, $v$, the efficiency, $\eta$, and the coefficient of performance of the refrigerator, $COP$, for various values of  barrier subdivision, $N$.
 Next, the energetics of the Brownian heat engine will be  explored as a function of model parameters both in the absence and   in the presence 
 of external force. 

\section{The efficiency and the velocity in the absence of external force}

In the absence of the external force $f=0$, the analytically obtained steady state current for $N=1$ is given by 
\begin{equation}
J={1\over 2\gamma(T_{h}+T_{c})}({U_{0}\over L_{0}})^2\left({1\over e^{U_{0}\over T_{h}}-1}-{1\over e^{U_{0}\over T_{c}}-1}\right) 
\end{equation} where $\gamma$ denotes the coefficient of friction of the Brownian particle. The magnitude of  the coefficient of friction of the Brownian particle $\gamma$ depends on  temperature of the viscous medium. As approximation, it is considered to be a constant. In Appendixes A and B,  the expressions for $F$, $G_{1}$, $G_{2},$ and  $H$ are given for $N=2$ and $N=3$, respectively. For $N\ge 4$, the expressions for   $F$, $G_{1}$, $G_{2},$ and  $H$  are lengthy and will not be presented in this work.

 When one omits the heat exchange  via kinetic energy (neglecting the term $1/2k_{B}(T_{h}-T_{c})$ in (3) and (4)),   the efficiency goes to Carnot efficiency $\eta_{c}$ at quasistatic limit.
 The quasistatic limit of the engine is obtained when $U_{0}$ goes to zero. For any $N$, we explore the efficiency at quasistatic limit and find that 
$  \displaystyle \lim _{{U_{0}\to 0}}{\eta }=(T_{h}-T_{c})/ T_{h},$ which is exactly equal to the efficiency of the  Carnot engine. 
\begin{figure}
\epsfig{file=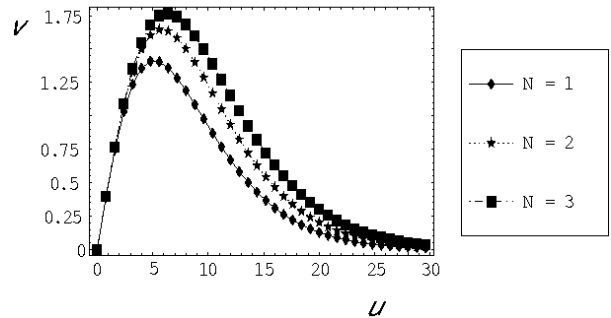,width=8cm}
\caption{ The drift  velocity  $v$ versus  the potential barrier $u$  for values  $\tau$=2 and  $\ell$=$1$. In the limit $u \to 0$ and $u \to \infty$, the velocity $v \to 0$. $v$  increases with the number of barrier subdivisions.  The velocity $v$ attains maximum values at a particular value of $u$. The potential $u$, at which the velocity of the particle is maximum, shifts to wards the left as the number of barrier subdivisions increase}
\end{figure}

\begin{figure}
\epsfig{file=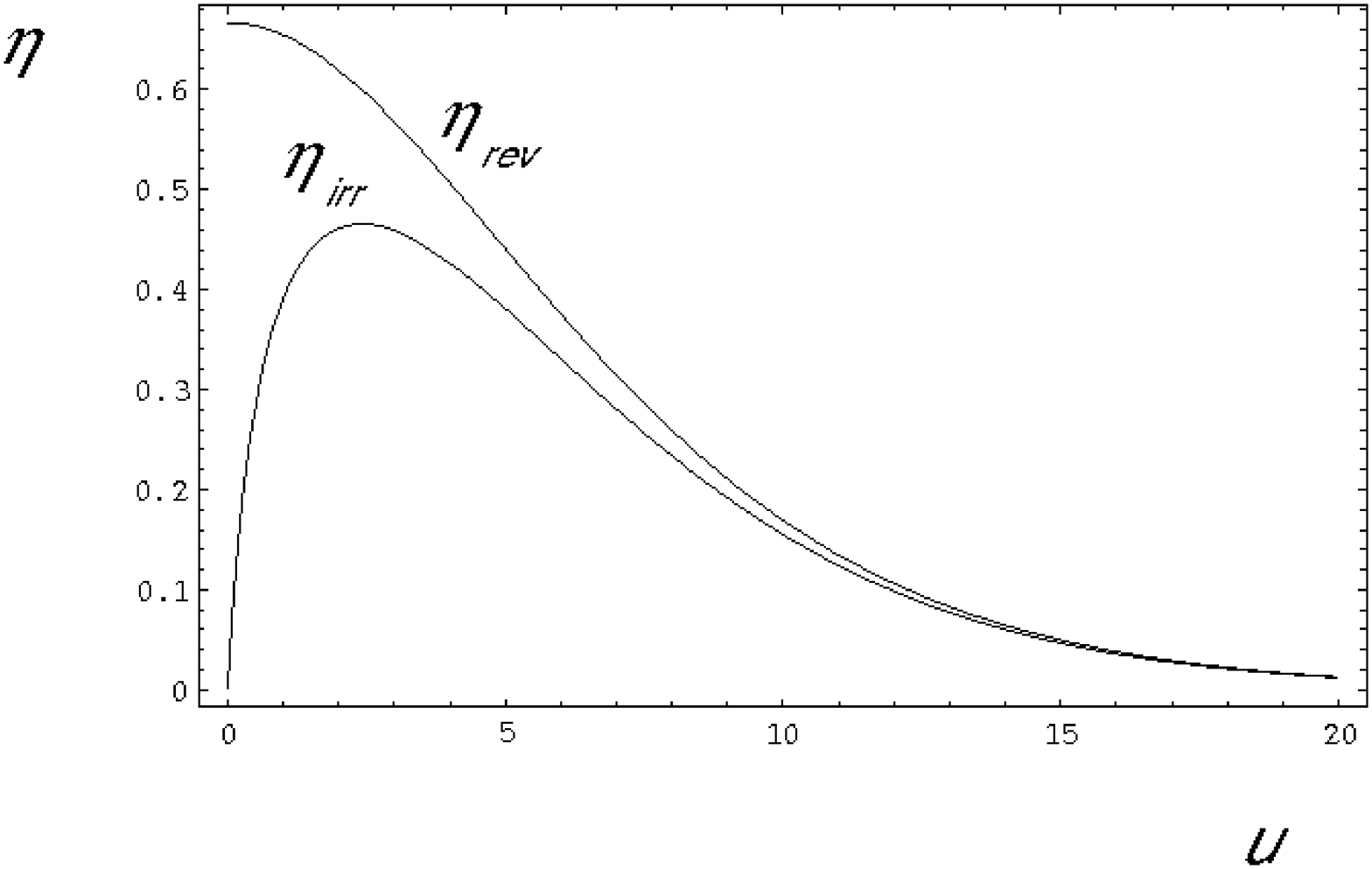,width=8cm}
\caption{ The efficiency  $\eta$ as a function of the barrier height   $u$ for parameter values of   $\tau$=2 and $\ell$=$1$.  In the limit $u \to 0$, $\eta_{rev} \to 2/3$ which is equal to the Carnot efficiency for  the given parameter values. In the limit $u \to 0$, $\eta_{irr} \to 0$. When $u \to \infty$, $\eta  \to 0$. }
\end{figure}

\begin{figure}
\epsfig{file=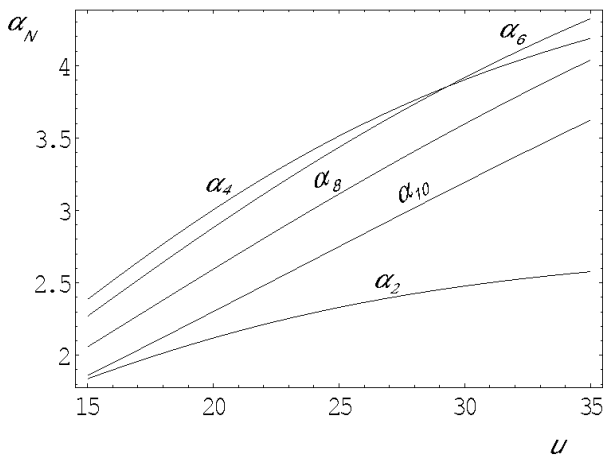,width=8cm}
\caption{ The plot of   $\alpha$ as a function of  $u$ for parameter values of   $\tau$=1.5 and $\ell$=$1$. $\alpha$   is increasing  function of   $u$.  For $u<29$,
$\alpha_{4}>\alpha_{2}$, $\alpha_{3}$,...$\alpha_{10}$. This implies $N=N_{op}=4$ is the optimal value  of barrier subdivision.  On other hand, when $u > 29$, $N=N_{op}=6$ is the optimal value  of barrier subdivision.}
\end{figure}

The heat flow via the kinetic energy significantly affects the efficiency of the Brownian heat engines.  Next, considering 
 the heat flow via the potential and the kinetic energies, we explore the thermodynamic property of the engine.
Let us introduce dimensionless parameters before exploring the dependence of the 
velocity and  the  efficiency on different values of   barrier subdivision $N$. We introduce 
scaled parameters: scaled length
 $\ell=1 $, scaled barrier height $ u=U_{0}/ k_BT_{c}$, scaled current $j=J/J_{0}$ where $J_{0}=k_BT_{c}/\gamma L_{0}^2$, scaled velocity $v=2j/\ell$ and scaled temperature  $\tau={T_h\over T_c}-1$. Here  $k_B$ denotes Boltzmann's constant. For simplicity, it is considered to be unity. We also introduce  dimensionless parameters  $\alpha_{i}=\eta_{i}/\eta_{1}$ ($i= 2,3$, $...$ $N$) where $\eta_{i}$ and  $\eta_{1}$ are the efficiencies  when $N=i$ and $N=1$, respectively.

The dependence of the steady state current or equivalently the drift velocity on the potential $u$ can be analyzed  by exploiting the analytically obtained steady state current (for instance, see Eq. (5) for the case $N=1$ and, the expressions shown in the Appendixes  A and B for the cases $N=2$ and 3, respectively).  The directional current is the result of spatial temperature difference along the ratchet potential. In the absence of the ratchet potential, the  average velocity of the particle  is zero, i.e.; 
the velocity  $v$ vanishes when $u \to 0$ (see Fig. 3). In the limit  $u \to  \infty$,  $j \to 0$ as the particle encounters  a difficulty of jumping the high potential barrier of the ratchet potential, see Fig. 3.  The velocity $v$ attains maximum value  at a particular value of $u$. Note that the engine operates with maximum power  at this particular value of $u$. 
 The potential $u$, at which the velocity of the particle is maximum, shifts to wards the right as the number of barrier subdivisions increase as it can be readily seen in Fig. 3.   Note that in the system we consider, the left  and  the right sides of the sawtooth potential are coupled with the hot and the cold  baths, respectively.  For such a system, positive velocity exhibits   that the net flux of the particle is from the hot to  the cold reservoirs and the engine operates only as a heat engine.   Figure 3 shows that the velocity is positive for any $N$.

For high potential barriers, the Brownian particle encounters a difficulty of jumping the sawtooth potential  when the  background temperature is weak. Subdividing the barrier along the reaction coordinates enables the Brownian particle to cross each small barrier with small thermal kicks and ultimately the particle crosses the high potential barrier within short period of time than the time taken by the particle  when it crosses the smooth potential barrier.  Hence subdividing the sawtooth potential enhances the drift velocity $v$ as depicted in Fig. 3. The possibility of enhancing the escape rate of a Brownian particle over sawtooth potential under specific conditions was envisaged  \cite{m1, m2}.
The analytical finding revealed that the escape rate of the Brownian particle is enhanced  for subdivided reaction coordinate which  qualitatively agrees with this work.
\begin{figure}
\epsfig{file=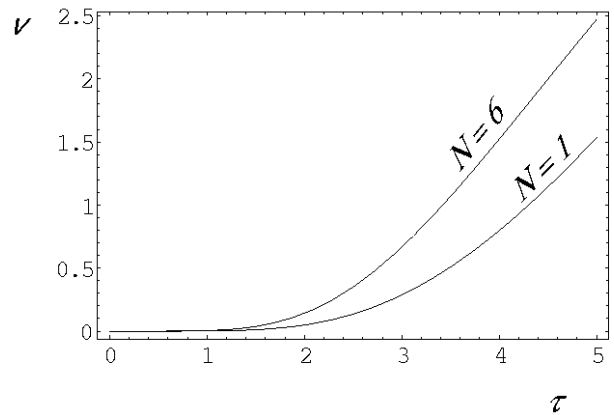,width=8cm}
\caption{ The drift velocity  $v$ versus $\tau$  for values of   $u$=24 and $\ell$=$1$. In the limit $\tau \to 0$ (since $T_{h}=T_{c}$, the system is in thermal equilibrium),  the  steady state current $j$ vanishes: $j \to 0$.
The drift velocity $v$ intensifies as $\tau$ and $N$ increase. }
\end{figure}

The analytically determined efficiency $\eta$ is plotted as a function of 
 $u$ in Fig. 4 for the case $N=1$. When one considers the heat flow via the potential energy,  in the limit $u \to 0$, $\eta_{rev} \to 2/3$  which is equal to the Carnot efficiency   for parameter values of $\tau$=2 and $\ell$=$1$.  On the other when we consider the heat flow both via the potential and the kinetic energies, $\eta_{irr} \to 0$ when  $u \to 0$  and  $u \to \infty$.
 This exhibits that quasistatic process may not be the best working condition for the Brownian heat engines which agrees with the  claim of Hondou and Sekimoto  \cite{ken}. In addition, $\eta_{irr}$ attains a maximum
value at finite value of $u$. The same figure depicts  that  $\eta_{irr} < \eta_{rev} $  and $\eta_{irr,rev} \to 0 $ when $u \to \infty$.

The plot of $\alpha$ as a function of $u$ is displayed in Fig. 5. The enhancement in the efficiency is high when $\tau$ is small. For $u<29$,
$\alpha_{4}>\alpha_{2}$, $\alpha_{3}$,...$\alpha_{10}$. This implies $N=N_{op}=4$ is the optimal value  of barrier subdivision, at which the efficiency (velocity) is maximum,  for a given parameter values. Note that the optimal barrier subdivision $N_{op}$ is sensitive to the choices of model parameters. For instance  when $u > 29$, $N=N_{op}=6$ is the optimal value  of barrier subdivision.

The net flux of the particle strictly depends on the temperature difference between the hot and the cold baths.  In the limit $\tau \to 0$ (since $T_{h}=T_{c}$, the system is in thermal equilibrium),  the  steady state current $j$ vanishes, i.e.; $j \to 0$ (see Eq. (5) for the case  $N=1$ and the expressions shown in the Appendixes  A and B for the cases  $N=2$ and 3, respectively). The unidirectional current $j$  is due to the non homogenous   temperature profile  along the ratchet potential as the particle in the hot bath can easily crosses the  potential barrier of the sawtooth potential  than the particle in the cold bath.    When the magnitude of the rescaled 
 temperature $\tau$ steps up,  the  tendency of the particle in the hot bath  to reach the top of the ratchet potential hill  increases  than  the  particle in the cold reservoir. This leads to an increase in the current $j$ or  the drift velocity $v$  as shown in Fig.  6. The same figure shows that the net flux of the particle intensifies  with $N$.    The plot of $\alpha$ as a function of $\tau$ is displayed in Fig. 7.
Significant enhancement of the efficiency  is observed when $\tau$ is small.

\begin{figure}
\epsfig{file=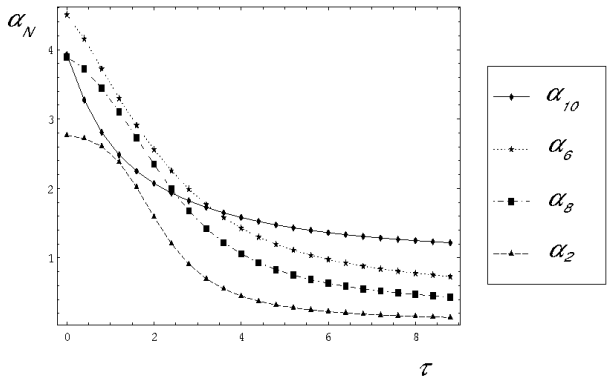,width=8cm}
\caption{ The plot of   $\alpha$ as a function of  $\tau$ for values of  $u$=24 and $\ell$=$1$. }
\end{figure}
   
   \section{The efficiency, the velocity and the performance of the refrigerator in the presence of external force}

In the presence of external force, the net flow of the particle depends on the 
magnitude of the external force. For large load, current reversal may occur and this  indicates that  the engine operates  not only as a heat engine but also as a refrigerator.    In the presence of  constant external force $f$,  similar to the previous section, the closed form expression for steady state  is given by $
     J={-F/( G_{1}G_{2} +HF)}.$
For $N=1$, the expressions for $F$, $G_{1}$, $G_{2},$ and  $H$ are given by
\begin{eqnarray}
  F&=& e^{a-b} - 1,\nonumber \\
  G_1&=&\frac{L_0}{aT_h}\left(1-e^{-a}\right) + \frac{L_0}{bT_c}e^{-a}\left(e^b - 1\right), \nonumber \\
  G_2&=&\frac{\gamma L_0}{a}\left(e^a - 1\right) + \frac{\gamma L_0}{b}e^a\left(1-e^{-b}\right).
  \end{eqnarray}
  On the other hand, $H=A+B+C$, where
  \begin{eqnarray}
  A&=&\frac{\gamma}{T_h} \left(\frac{L_0}{a}\right)^2 (a + e^{-a}-1), \nonumber \\
  B&=&\frac{\gamma L_0L_0}{abT_c} (1-e^{-a})(e^b-1), \nonumber \\
  C&=& \frac{\gamma}{T_c}\left(\frac{L_0}{b}\right)^2(e^b-1-b).
  \end{eqnarray}
  Here  $a = (U_0 + f L_0)/T_h$ and $b = (U_0 - f L_0)/T_c$.
The expressions for $N \ge 2$ are lengthy and will not be presented in this work. 
The drift velocity $v$ is related to the steady state current and it is given by   $ v=2JL_{0}$.
Introducing additional rescaled  parameter: $\lambda=fL_{1}/T_{c}$,  we study how the velocity, the efficiency and the coefficient of performance of the refrigerator  behave as $N$ varies. 

\begin{figure}
\epsfig{file=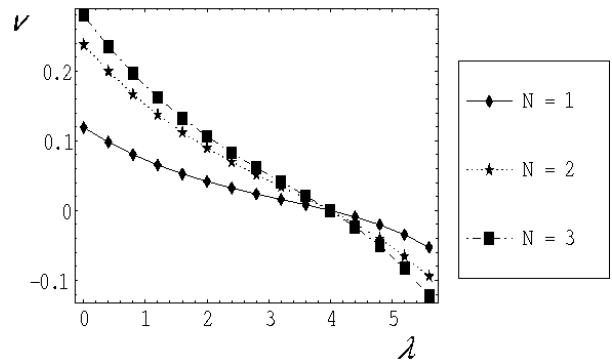,width=8cm} 
\caption{ The drift  velocity  $v$ versus  the rescaled load $\lambda$ for values    $u=12$, $\ell=1$ and $\tau=1$.  For $\lambda<4$, the net particle flow is from the hot to the cold reservoirs. The drift velocity  $v=0$ when   $\lambda=4$. Current  inversion occurs when   $\lambda>4$. }
\end{figure}
Figure 8 presents the plot of the velocity $v$ versus rescaled load $\lambda$ for $N=1$, $2$ and $3$. As shown in the figure for $\lambda < 4$, the load is not strong enough to reverse the direction of the net flux of the particle, i.e., 
the net flow of the  particle is from the hot  to the cold reservoirs and hence the model works as a heat engine. On other hand for $\lambda > 4$, the current becomes negative and the model acts as a refrigerator. The particle velocity is zero at $\lambda =4$. In general, for any number of barrier subdivisions  $N$  the velocity $v=0$, when the load is 
\begin{equation}
f_0={\tau U_{0}\over (\tau+2)L_{0}}.
\end{equation}
When one omits the heat exchange  via kinetic energy (neglecting the term $1/2k_{B}(T_{h}-T_{c})$ in (3) and (4)), 
for any $N$,    in the  quasistatic limit $v^+\to 0$,  the efficiency is equal to Carnot efficiency:
\begin{equation}
 \displaystyle \lim _{v^+\to 0}{\eta_{C} }={T_{h}-T_{c}\over T_{h}}
\end{equation}
 and in the quasistatic limit $v^-\to 0$, the coefficient of performance of the refrigerator  is equal to Carnot refrigerator:
\begin{equation}
 \displaystyle \lim _{v^-\to 0}{P_{ref} }={T_{c}\over T_{h}-T_{c}}.
\end{equation}
We further investigate the thermodynamic property of the engine by including the heat exchange via the kinetic  and the potential energies.  At a  quasistatic limit, for any $N$,   the efficiency takes a simple form:
\begin{equation}
\displaystyle \lim _{v^+\to 0}\eta_{irr}^{*} ={T_{h}-T_{c}\over T_{h} }\delta=\eta_{c}\delta
\end{equation}
where 
\begin{equation}
\delta={2U_{0}\over {2U_{0}T_{h}\over T_{c}}+0.5{T_{h}^2\over T_{c}}-0.5T_{c}}.
\end{equation}
Note that  $\delta<1$ when $T_{h}>T_{c}$. This exhibits  that the efficiency of the engine never approaches  the Carnot efficiency at quasistatic limit as there is irreversible heat transfer via the kinetic energy  from the hot to the cold heat baths. The unattainability of the Carnot efficiency, at a quasistatic limit, was reported in the work \cite{ken} by solving the Kramers equation at the stall force. Exploiting the  analytical expressions (9) and (11), one can explore   how $\eta_{irr}^{*}$  and $\eta_{C}$ behave as a function of $\tau$ as displayed in Fig. 9. The figure shows that 
 in the  limit $T_{h} \to T_{c}$, $\eta_{irr}^{*} \to \eta_{C}$  while  $\eta_{irr}^{*} \ll \eta_{C}$ when  $T_{c}\ll T_{h}$.  
This exhibits that the heat transfer via the kinetic energy is considerable  when $\tau$  steps up.  

\begin{figure}
\epsfig{file=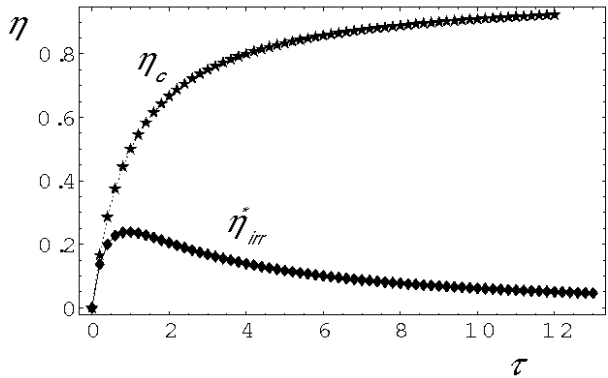,width=8cm} \caption{The efficiency  $\eta$ versus $\tau$  when the engine works quasistatically.
$\eta_{C}$ denotes the Carnot efficiency  while $\eta_{irr}^{*}$  denotes the efficiency of  the irreversible heat  engine when it operates quasistatically for fixed $u=2$. }
\end{figure}

 \begin{figure}
\epsfig{file=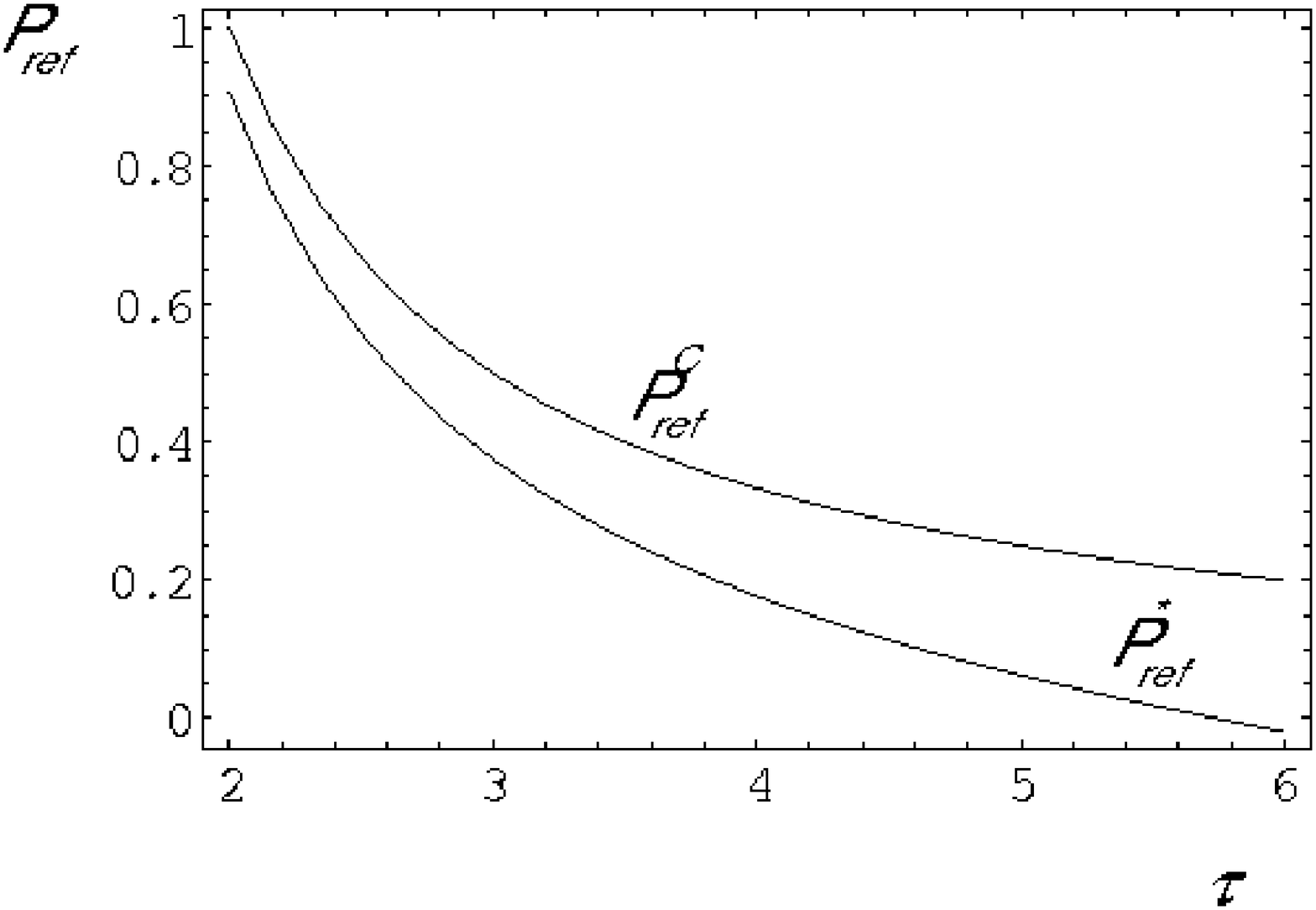,width=8cm} \caption{The coefficient of performance of the refrigerator  $P_{ref}$ versus $\tau$ when the heat engine works quasistatically.  $P_{rec}^{C}$ denotes the Carnot refrigerator  while $P_{irr}^{*}$  denotes $P_{ref}$  of   the irreversible heat  engine when it operates quasistatically for fixed $u=8$. }
\end{figure}

In the limit $v^-\to 0$, the coefficient of performance of the refrigerator converges to 
\begin{equation}
 \displaystyle \lim _{v^-\to 0}{P_{ref} }={T_{c}\over T_{h}-T_{c}}\Delta.
\end{equation}
where 
\begin{equation}
\Delta ={U_{0}{T_{c}\over T_{h}}-0.25{T_{h}^2\over T_{c}}+0.25T_{c}\over U_{0}}.
\end{equation} 
When $T_{h}>T_{c}$  and within the region where the model works as a refrigerator, $\Delta <1$. This reveals that  
the coefficient of performance of the refrigerator $P_{ref}$ is always less than the Carnot refrigerator at a quasistatic limit.   The quasistatic behavior of the engine can be explored   by exploiting the analytic expressions (10) and (13) as shown in Fig. 10. The figure depicts that 
 $P_{ref}^{*} < P_{ref}^{C}$ and, in the  limit $T_{h} \to T_{c}$, $P_{ref}^{*} \to P_{ref}^{C}$.

\begin{figure}
\epsfig{file=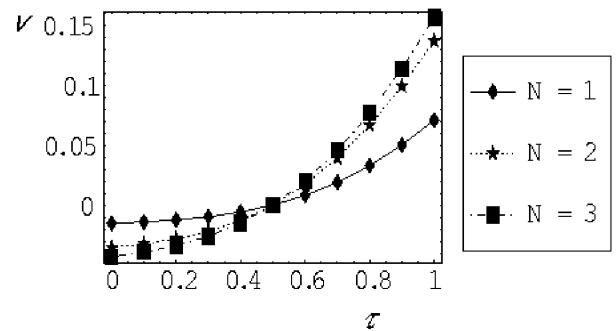,width=8cm}
\caption{ The drift  velocity  $v$  as a function of    rescaled temperature $\tau$ for fixed parameters $u=10$, $\ell=1$ and $\lambda=2.0$. For $\tau< 0.5$, the velocity $v$ takes  negative values and the engine works as a refrigerator within this region. At $\tau=0.5$, the velocity $v=0$ while when $\tau>0.5$, the net particle current is from the hot to the cold reservoirs.  As shown in the figure, $|v|$ intensifies  as the number of barrier subdivisions  increase.}
\end{figure}

\begin{figure}
\epsfig{file=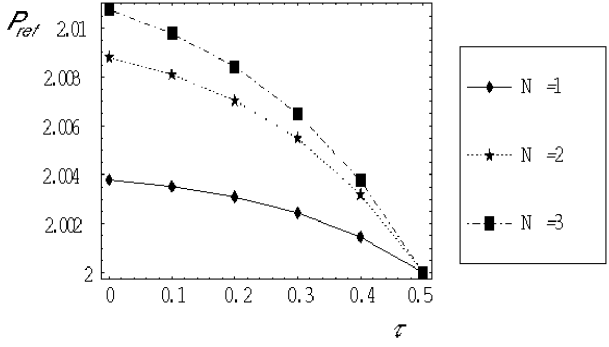,width=8cm} \caption{The coefficient of performance of the refrigerator  $P_{ref}$ versus $\tau$ (for reversible case) for fixed   $u=10$, $\ell=1$ and $\lambda=2.0$. $P_{ref}$ is a decreasing function of $\tau$ and it  rises up with  $N$. In the limit $\tau \to 0.5$,   $P_{ref} \to 2$ which is equal to the Carnot refrigerator for  the given parameter values. }
\end{figure}

We  next explore how  $v$ and $P_{ref}$ behave as a function of $\tau$  for  $N=1$, 2 and 3.  The dependence of the velocity $v$ on the rescaled temperature $\tau$ is demonstrated in Fig. 11  for $N=$1, 2 and 3. As shown in the Fig. 11,  when $\tau<0.5$, the load is strong enough to reverse the net particle flow and  the velocity is negative. On other hand,  when $\tau=0.5$, the temperature renormalizes the effect of the load and $v=0$. For $\tau>0.5$, the temperature gains strength to overcome the load and hence the current is positive in this region.
Within the region where the model works as a heat engine, $v$ strengthens  with  $N$ and $\tau$.  In the region where  the model works as a refrigerator, $|v|$ intensifies   as   $N$ increases as shown in Fig. 11. This is because subdividing  the sawtooth potential  into small barriers enables the  Brownian particle to   cross these small barriers at small thermal kicks. On the other hand, when $\tau$ is small, the background thermal kick is weak for the particle to cross the smooth sawtooth potential barrier. 

Figure 12 depicts the plot of $P_{ref}$ versus $\tau$.  The coefficient of performance of the  refrigerator $P_{ref}$ is a decreasing function of $\tau$ and it attends a maximum value when $N$ steps up. When the rescaled temperature $\tau$ increases, 
 $P_{ref}$ declines  towards the Carnot refrigerator.

\section{Summary and conclusion}

In this work, we consider a model of Brownian heat engine. The dependence of the velocity, the efficiency and the coefficient of performance of the refrigerator is investigated for different number of barrier subdivisions, $N$. We show that  the velocity and the efficiency attain optimum values  at a particular value of barrier subdivision  $N$. In the presence of external load we find that  the velocity, the efficiency and the coefficient of  performance of  the refrigerator attain maximum values when 
 the sawtooth potential is subdivided  into series of smaller connected barrier series. 

Considering the heat exchange  via the potential and the kinetic energies, we show that  Carnot efficiency is unachievable for 
Brownian heat engines when the engines work quasistatically. Quasistatic consideration for the Brownian heat engines also reveals that 
  the coefficient of performance of the refrigerator is always less than the Carnot refrigerator.

In this work, considering an exactly solvable model, we explore the energetics of a Brownian heat engine not only at quasistatic limit but also at any finite time. This theoretical work suggests that the performance of the heat engine can be improved by subdividing the sawtooth potential into series of small barrier steps systematically by considering physically reasonable parameterization.

   \section*{Acknowledgements}
I would like to thank Mulugeta Bekele for the interesting discussions and, for his very helpful comments and suggestions. I would like also to thank him for his careful and critical reading of this manuscript. It is my pleasure to thank  Hsuan-Yi Chen for the interesting discussions and for providing a wonderful research environment.

\appendix
  \begin{widetext}
\renewcommand{\theequation}{A\arabic{equation}}
  % redefine the command that creates the equation no.
  \setcounter{equation}{0}  % reset counter
\section*{Appendix A}
In this Appendix we will give the expressions for $F$, $G_{1}$, $G_{2}$ and   $H$ which define the value of the steady state current, $J^{L}$, for zero external load case when $N=2$.
\begin{eqnarray}
F&=&e^{{-U_{0}\over T_{c}}+ {U_{0}\over T_{h}}}-1,\\
G_{1}&=&{3L_{0}\over 5U_{0}}e^{-U_{0}\over T_{h}}[-2-2e^{ U_{0}\over 3T_{C}}+2e^{2U_{0}\over 3T_{C}}+e^{U_{0}\over T_{C}}-2e^{U_{0}\over 3T_{h}}+2e^{2U_{0}\over 3T_{h}}]+ {3L_{0}\over 5U_{0}},\\
G_{2}&=& {L_{0}\gamma\over  5U_{0}}(-3e^{ ({1\over T_{h}}-{1\over T_{C}})U_{0}}T_{C}-3T_{h}-6e^{ U_{0}\over 3T_{h}}T_{h}+6e^{2U_{0}\over 3T_{h}}T_{h})+ \nonumber \\ & & {3L_{0}\gamma \over 5U_{0}}e^{{-2U_{0}\over 3T_{C}}+{U_{0}\over T_{h}}}[-6T_{C}+6e^{U_{0}\over 3T_{C}}T_{C}+3e^{2U_{0}\over 3T_{c}}(T_{C}+T_{h})],\\
H^{L}&=&T_{1}(T_{2}+T_{3}+T_{4}+T_{5}+T_{6}), \\
  T_{1}&=&{L_{0}^{2}\gamma\over 25T_{h}U_{0}^2} e^{-{1\over 3}({1\over T_{C}}+{3\over T_{h}})U_{0}},\\
   T_{2}&=&36e^{U_{0}\over T_{h}}T_{C}T_{h}+18e^{U_{0}\over 3T_{C}}T_{h}^2+18e^{2U_{0}\over 3T_{C}}T_{h}^2-18e^{U_{0}\over 3_{C}}T_{h}^2-9e^{4U_{0}\over 3T_{C}}T_{h}^2, \\
 T_{3}&=&-36e^{{1\over 3} ({3\over T_{C}}+{1\over T_{h}})U_{0}}T_{h}^2+45e^{ (T_{C}+T_{h})U_{0}\over 3T_{C}T{h}}T_{h}^2-36e^{2(T_{C}+T_{h})U_{0}\over 3T_{C}T{h}}T_{h}^2 ,\\
T_{4}&=&54e^{2(T_{C}+T_{h})U_{0}\over3T_{C}T{h}}T_{h}^2+36e^{(4T_{C}+T_{h})U_{0}\over 3T_{C}T{h}}T_{h}^2+18e^{2(T_{C}+2T_{h})U_{0}\over 3T_{C}T{h}}T_{h}^2+ \nonumber \\ & &  
36e^{(T_{C}+2T_{h})U_{0}\over 3T_{C}T{h}}T_{h}^2,\\
T_{5}&=&-18e^{(T_{C}+4T_{h})U_{0}\over 3T_{C}T{h}}T_{h}^2+36e^{{ U_{0}\over T_{C}}+{2U_{0}\over 3T_{h}}}T_{h}^2+9e^{{4U_{0}\over 3T_{C}}+{ U_{0}\over T_{h}}}(T_{h}(T_{h}+T_{C}),\\
T_{6}&=&18e^{ (T_{C}+T_{h})U_{0}\over T_{C}T{h}}T_{h}(2T_{C}+T_{h})-18e^{{2U_{0}\over 3T_{C}}+{ U_{0}\over T_{h}}}T_{h}(2T_{C}+T_{h})+ \nonumber \\ & & e^{{U_{0}\over 3T_{C}}+{U_{0}\over T_{h}}}(-45T_{h}(T_{C}+T_{h})-6T_{h}U_{0}+2U_{0}^2).
    \end{eqnarray}

\renewcommand{\theequation}{B\arabic{equation}}
  % redefine the command that creates the equation no.
  \setcounter{equation}{0}  % reset counter
\section*{Appendix B}
In this Appendix we will give the expressions for $F$, $G_{1}$, $G_{2}$ and   $H$ which define the value of the steady state current, $J$, for zero external load case when $N=3$.
\begin{eqnarray}
F& = &e^{{-U_{0}\over T_{c}}+{ U_{0}\over T_{h}}}-1,\\
G_{1}& =&{L_{0}e^{-U_{0}\over T_{h}}\over 2U_{0}}(-2-2 e^{U_{0}\over 4T_{C}}+ 2e^{3U_{0}\over 4T_{C}}+ e^{U_{0}\over T_{C}}- 2e^{U_{0}\over 4T_{h}}+2 e^{3U_{0}\over 4T_{h}}+ e^{U_{0}\over T_{h}}),\\
G_{2}& =&{-L_{0}\gamma \over 2U_{0}}(e^{U_{0} ({1\over T_{h}}-{1\over T_{C}})}T_{C}+2e^{{ U_{0}\over T_{h}}-{ 3U_{0}\over 4T_{C}}}T_{C}-2 e^{{ U_{0}\over T_{h}}-{ U_{0}\over 4T_{C}}}T_{C})+\nonumber \\ & &
{-L_{0}\gamma \over 2U_{0}}(T_{h}+2e^{ U_{0}\over 4T_{h}}T_{h}-2e^{3U_{0}\over 4T_{h}}T_{h}-e^{ U_{0}\over T_{h}}(T_{h}+T_{C})),\\
H&=&T_{1}+T_{2}(T_{3}+T_{4}),\\
T_{1}& =&{L_{0}^{2}\gamma T_{C}\over 4U_{0}^2}(-5+8e^{- U_{0}\over 4T_{C}}-12 e^{ U_{0}\over 4T_{C}}+4 e^{ U_{0}\over 2T_{C}}+4 e^{3U_{0}\over 4T_{C}}+ e^{ U_{0}\over T_{C}}),\\
T_{2}&=&{L_{0}^{2}\gamma T_{h}\over 4U_{0}^2}(e^{ U_{0}\over 2T_{h}}-1)e^{- U_{0}\over T_{h}},\\
T_{3}& =& (-2-2e^{ U_{0}\over 4T_{C}}+2 e^{3U_{0}\over 4T_{C}}+e^{ U_{0}\over T_{C}}+e^{0.5U_{0}({2\over T_{c}}+{1\over T_{h}})},\\
T_{4}& =&2e^{0.25U_{0}({1\over T_{h}}+{4\over T_{C}})}+ 2e^{0.25U_{0}({2\over T_{h}}+{3\over T_{C}})}- 6e^{ U_{0}\over 4T_{h}}- 6e^{ U_{0}\over 2T_{h}}+8e^{3U_{0}\over 4T_{h}},\\
T_{5}& =& -4e^{ U_{0}(T_{h}+T_{C})\over 4 T_{h}T_{C}}-2e^{ U_{0}(T_{h}+2T_{C})\over 4 T_{h}T_{C}}+4e^{ U_{0}(3T_{h}+T_{C})\over 4 T_{h}T_{C}}.\\
 \end{eqnarray}
\end{widetext}


\begin{thebibliography}{16}
\bibitem{ph} R.D. Astumian, P. Hanggi, Phys. Today {\bf 55}, 33 (2002).
\bibitem{r} P. Reimann, Phys. Rep. {\bf 361}, 57 (2002).
\bibitem{ken} Tsuyoshi Hondou and Ken Sekimoto, Phys. Rev. E {\bf 62}, 6021 (2000).
\bibitem{gom} A. Gomez-Marin and  J.M. Sancho, Phys. Rev. E {\bf 74}, 062102 (2006).
\bibitem{butt87} B\"{u}ttiker M., \emph{Z. Phys. B} {\bf 68}, 161 (1987).
\bibitem{kampen88} Van Kampen N. G., \emph{IBM J.Res. Dev.} {\bf 32}, 107 (1988).
\bibitem{land88} Landauer R., \emph{J. Stat. Phys.} {\bf 53}, 233 (1988).
\bibitem{miki3} Miki Matsuo and Shin-ichi Sasa, \emph{Physica A} {\bf 276}, 188 (1999).
\bibitem{Aus1} Der\`enyi I. and Astumian R. D., \emph{Phys. Rev. E} {\bf 59}, R6219 (1999).
\bibitem{Ast2} Der\`enyi I., Bier M. and Astumian R. D., \emph{Phys. Rev. Lett} {\bf 83}, 903 (1999).
\bibitem{bq} B.Q. Ai, H.Z. Xie, D.H. Wen, X.M. Liu, L.G. Liu, Eur. Phys. J. B {\bf 48}, 101 (2005) 
\bibitem{mesfin1} Mesfin Asfaw and Mulugeta Bekele, Eur. Phys. J. B {\bf 38},  457 (2004).
\bibitem{mesfin2} Mesfin Asfaw and Mulugeta Bekele, Phys. Rev. E {\bf 72}, 056109 (2005).
\bibitem{mesfin3} Mesfin Asfaw and Mulugeta Bekele, Physica  A {\bf 384}, 346 (2007).
\bibitem{mesfin4} Mesfin Asfaw, Senior B.Sc. thesis, Addis Ababa University, Addis Ababa, (1999), (unpublished).
\bibitem{m1} Mulugeta Bekele, G. Ananthakrishna and N. Kumar, Pramana - J. Phys. {\bf 46}, 403 (1996).
\bibitem{m2} Mulugeta Bekele, G. Ananthakrishna and N. Kumar, Physica  A {\bf 270}, 149 (1999). 
\bibitem{g1}I. Goldhirsch and Y. Gefen, Phys. Rev. A {\bf 33}, 2583 (1986).
\bibitem{g2}I. Goldhirsch and Y. Gefen, Phys. Rev. A {\bf 35}, 1317 (1987).
\bibitem{san} J.M. Sancho, M. San Miguel, D. D\"urr, J. Stat. Phys. {\bf 28}, 291 (1982).
\end{thebibliography}
\end{document}